# Improvement of a 1-D Fluid Probe Model for Mach Probe Measurements

P. Peleman[1], S. Jachmich[2], M. Van Schoor[2], G. Van Oost[1]

[1] Ghent University – Department of Applied Physics, B-9000 Ghent, Belgium
[2] Laboratory for Plasma Physics, Ecole Royale Militaire/Koninklijke Militaire School, Partner Tril. Euregio Cluster, Euratom association "Belgian State", B-1000 Brussels, Belgium

## 1. Abstract

In this paper we show how a two dimensional fluid model can be used to interpret data obtained from an inclined Mach-probe or a Gundestrup probe. We use an analytical approximation of the solution of the differential equations describing the relation between the plasma flow and the measured ion saturation currents at the probe's surface. The parameters of this approximate analytical solution are determined by comparison with the exact numerical solution of the equations. In this way we are able to measure the parallel as well as the perpendicular Mach numbers over the whole parameter range with a minimum accuracy of 90%.

## 1. Introduction

Mach probes are a common diagnostic to measure the flows and electric field profiles in the edge of fusion machines. The Mach numbers ($M_{//,\perp}$) of the parallel and perpendicular (in the magnetic surface, but perpendicular to the magnetic field) flow of the unperturbed plasma are derived from the ratio R of the up- and downstream ion saturation currents. To derive $M_{//}$, Hutchinson [1] has developed a 1-D fluid probe model, which has been extended by Van Goubergen [2] to study the influence of $M_\perp$ on the ratio R of the up- and downstream current, when the inclination angle $\theta$ of the probe surface with respect to the magnetic field is changed. These models essentially relate the ion saturation currents measured at the probes surfaces to the Mach numbers of the flow of the plasma not perturbed by the probe via a set of coupled differential equations. The numerical solutions of these equations can be approximated by an analytical function $\ln(R) = c[M_{//} + M_\perp \cot(\theta)]$, were c is a constant equal to 2.3 [3]. However, a comparison with the numerical solution of the differential equation shows that c depends on $M_{//}, M_\perp$ and $\theta$, resulting in an underestimation of the Mach numbers when the analytical model with constant $c$ is used. For values up to $M_{//,\perp} = 0.6$ the error made is rather small but increases to 25% for larger values as for example encountered in biasing experiments [4,5,6]. Based on the numerical (exact) solution of the differential equations we developed an expression for $c$ in which the dependency on $M_{//}, M_\perp$ and $\theta$ has been taken into account. This technique allows us to drastically reduce the errors. In the following section we describe the fluid model and we show the consequences of the use of a constant $c$. We then introduce the proposed function for $c$ and quantify the improvement.

## 2. Description of a Mach probe by a 1-D fluid model

Hutchinson's model starts from the continuity equation and the parallel ion momentum equation [1]. Combination of these equations results in a 1-D model that relates the density and the parallel Mach number at infinity ($n, M_{//,\infty}$) to the density at the probe surface ($n_s$) which can be measured via the ion saturation current given by $I_{sat} = n_s c_s A$, where A is the surface of the collector and $c_s$ the sound speed of the ions. To measure the perpendicular flow one has to incline the surface of the collectors with respect to the flow and extend the model as has been done by Van Goubergen [2]. The resulting equations are:

$$\frac{\partial n}{\partial //} = \frac{\left(M_{//} - \frac{M_\perp}{\tan(\theta)}\right)(1-n) - (M_{//,\infty} - M_{//})}{\left(M_{//} - \frac{M_\perp}{\tan(\theta)}\right)^2 - 1} \qquad \frac{\partial M_{//}}{\partial //} = \frac{-(1-n) + \left(M_{//} - \frac{M_\perp}{\tan(\theta)}\right)(M_{//,\infty} - M_{//})}{n\left[\left(M_{//} - \frac{M_\perp}{\tan(\theta)}\right)^2 - 1\right]}$$

(1) (2)

The angle $\theta$ is the angle between the magnetic field and the collectors as shown in figure 1. All the other symbols indicate dimensionless quantities and are defined in [2].

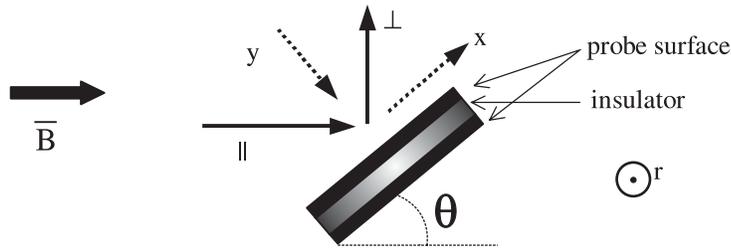

Figure 1: Mach probe geometry showing the parallel $(//)$, perpendicular $(\perp)$ and radial $(r)$ directions. The inclination angle of the collectors, with respect to the magnetic field, is defined by $\theta$

The unperturbed plasma is thus described by the parallel Mach number $M_{//,\infty}$ and a normalized density $n = 1$. With these starting values we solve equations (1) and (2) numerically and obtain the spatial variation of the density and parallel Mach number in the pre-sheath as shown in figure 2 and 3. The non-dimensional parallel distance is chosen such that $// = -\infty$ defines the unperturbed plasma and $// = 0$ the Magnetic Pre-sheath Entrance (MPSE) where the Debye sheath starts, defined by the Bohm boundary condition: $M_{//,MPSE} = \frac{M_\perp}{\tan(\theta)} \pm 1$. (3)

This condition is a result from the singularity of the denominators of equation (1) and (2). Figures 2 and 3 show three cases ($\theta = 50°, 90°$ and $130°$) for given values of $M_{//,\infty}$ and $M_\perp$. We plotted the evolution of the density (figure 2) and of the parallel Mach number (figure 3), both for the up- and downstream collectors. We define the upstream collector as the one which faces the flow vector in the direction of the magnetic field. When $\theta = 90°$, the system is insensitive to perpendicular flow and the ions reach the sound speed at the MPSE. When the probe is inclined, perpendicular flow is measured, and the parallel Mach number at the MPSE has to adapt itself to a value imposed by equation (3). Figure 2 shows that, due to the conservation of particles, the ion density in the pre-sheath must decrease when the ions accelerate towards the MPSE.

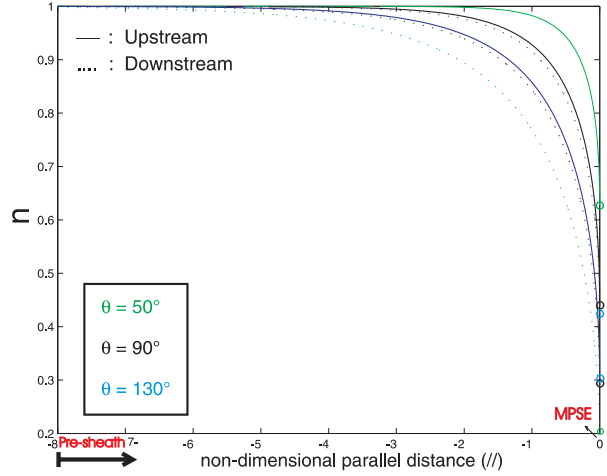
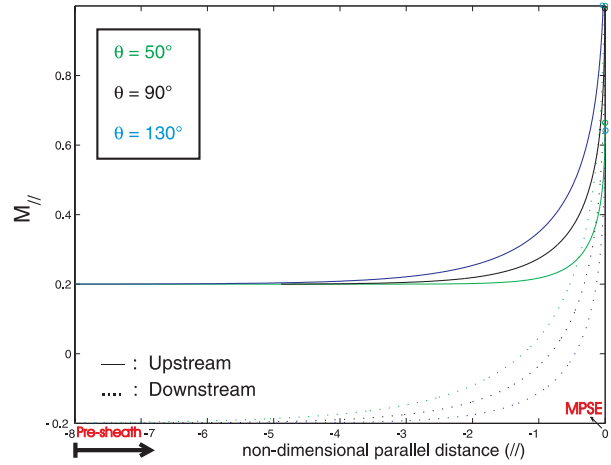

Figure 2: The spatial variation of the density in the pre-sheath for $|M_{//,\infty}| = 0.2$ and $|M_\perp| = 0.4$

Figure 3: The spatial variation of the parallel Mach number in the pre-sheath for $|M_{//,\infty}| = 0.2$ and $|M_\perp| = 0.4$

Dividing equation (1) by (2) immediately gives the evolution of the density as a function of the parallel Mach number:

$$\frac{\partial n}{\partial M_{//}} = n \frac{\left(M_{//} - \frac{M_\perp}{\tan(\theta)}\right)(1-n) - \left(M_{//,\infty} - M_{//}\right)}{-(1-n) + \left[\left(M_{//} - \frac{M_\perp}{\tan(\theta)}\right)\left(M_{//,\infty} - M_{//}\right)\right]} \qquad (4)$$

A solution for a given $|M_{//,\infty}| = 0.2$ and $|M_\perp| = 0.4$ is shown in figure 4. The value at density n=1 defines the parallel Mach number of the unperturbed plasma. The curves end at the MPSE, hereby defining the values of $M_{//,MPSE}$ and $n_{sh}$. If we apply this procedure in the range $-1 \leq M_{//,\infty} \leq 1$ for a given $M_\perp$ and $\theta$ and retain the associated sheath density at the MPSE, a relation between $M_{//,\infty}$ and $n_{sh}$ is obtained. An example is shown in figure 5 for $M_\perp = 0.4$ and four different values for $\theta$.

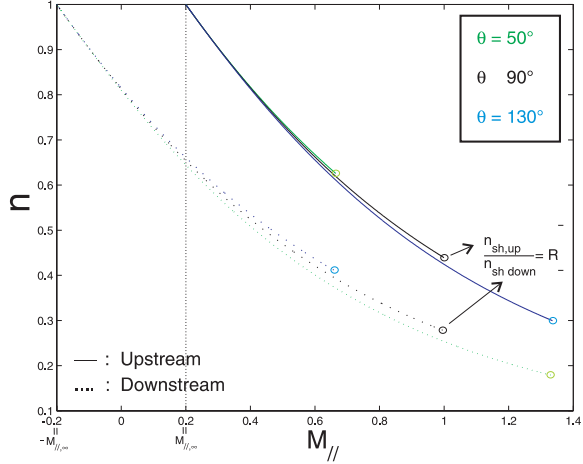
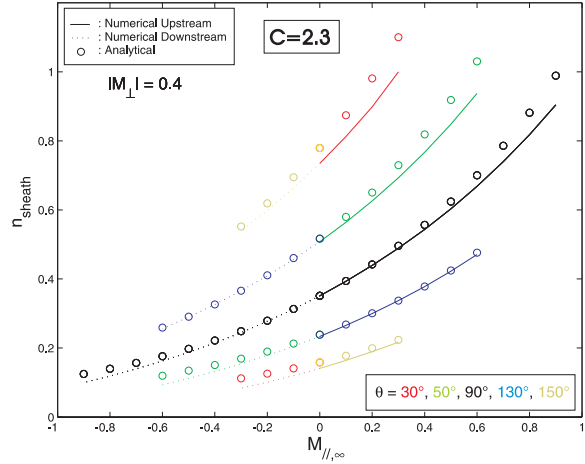

Figure 4: The normalized density as a function of $M_{//}$ in the pre-sheath for $|M_{//,\infty}| = 0.2$ and $|M_\perp| = 0.4$

Figure 5: The numerical and approximated analytical solutions of $n_{sheath}$ versus $M_{//,\infty}$

For the experiment, the ratio $R = \dfrac{I_{sat,up}}{I_{sat,down}}$ is important. With the numerical results we can determine this ratio via $R = \dfrac{n_{sh,up}}{n_{sh,down}}$. An approximate analytical solution of equation (4) for the density at the pre-sheath entrance, was proposed by Hutchinson and extended by Van Goubergen:

$$n_{sh_{down}^{up}} = \exp\left[c_{_{down}^{up}}\left(\pm|M_{//,\infty}|\pm\dfrac{|M_\perp|}{\tan(\theta)}\right)-c_0\right]. \qquad (5)$$

The values of $c_{_{down}^{up}}$ and $c_0$ can be determined by taking values of $n_{sh}$ produced by the numerical solution of equation (4). Setting $M_{//,\infty} = 0$ and $\theta = 90°$ in equation (5), one finds $c_0 = -\ln(n_{sh}) \approx 1.05$, while $c_{_{down}^{up}} = \dfrac{\ln\left(n_{sh_{down}^{up}}\right)+c_0}{\pm|M_{//,\infty}|}$ depends on $M_{//,\infty}$. This choice makes the fit between the approximate analytical solution and the exact numerical solution rather good for small values of $M_{//,\infty}$ and $\theta$ close to 90°. The solution however diverges for higher values of $M_{//}$ (as shown in figure 6) and the error can no longer be neglected. For the ratio R we get $\ln(R) = c\left[|M_{//,\infty}|+|M_\perp|\cot(\theta)\right]$ with $c = c_{up} + c_{down}$. The dependency of $c$ on $M_{//,\infty}$, $M_\perp$ and $\theta$ was found to be weak. In the past a constant value of $c = 2.3$ was used [3] and the disagreement with the numerical exact solutions, when not taking into account any dependency, was ignored. However in figures 5 and 6 one can see that, under certain conditions, the error can no longer be neglected. For example, for a given $M_\perp = 0.5$ and $M_{//,\infty} = 0.4$ (figure 6, dashed vertical line) the results diverge from the numerical ones for bigger inclination angles of the probe. On the other hand when $\theta$ is kept constant the error builds up with growing parallel flow. The latter effect is also shown in figure 7 where we plot the perpendicular Mach number of the approximated versus numerical solution for a constant $\theta = 100°$. We conclude that the error increases with growing parallel Mach number. Furthermore, for these settings, the weak dependency on $M_\perp$ is demonstrated by a nearly constant slope of the curves.

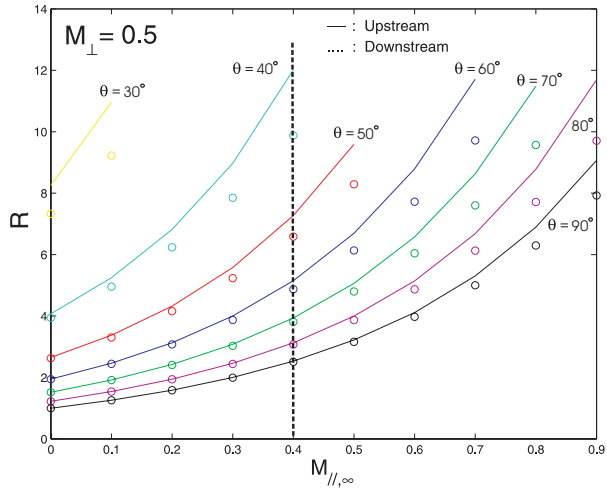 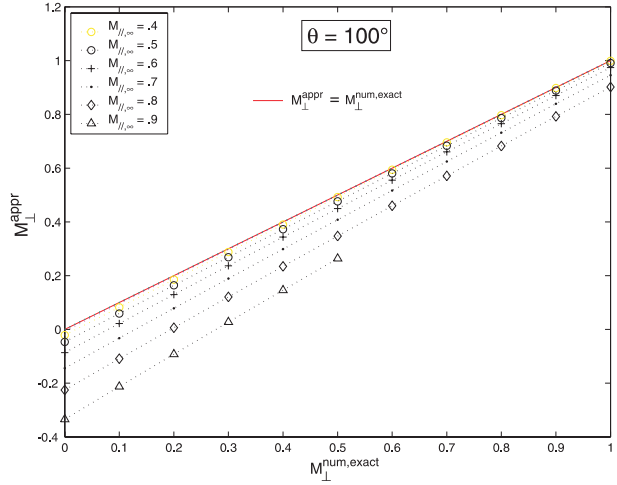

Figure 6: The ratio R as a function of $M_{//,\infty}$ for different inclination angles of the probe at a fixed $M_\perp = 0.5$

Figure 7: The approximated $M_\perp^{appr.}$ versus the numerical $M_\perp^{num}$. The red solid line represents the case for which $M_\perp^{appr.}$ is equal to $M_\perp^{num}$.

Therefore, in the following we present a better definition for c, which minimizes the error between the approximated analytical solutions and the exact numerical solution and so the underestimation of the flows.

## 3. Improvement of the approximated expression

Basically we will investigate the possibility to derive an analytical expression for $c(M_{//,\infty}, M_\perp, \theta)$ over the complete parameter range, $0 \leq |M_{//,\perp}| \leq 1$ and $0 \leq \theta \leq 180$. Instead of using a constant value for $c$, we assume the following expression $c(M_{//,\infty}, M_\perp, \theta) = c_1(M_{//,\infty}) c_2(M_\perp, \theta)$ which reduces the parameter study to a 2-D problem. Based on the following results this assumption is justified. The logarithm of R is calculated from the numerical solutions of the differential equation for a set of data of $M_{//,\infty}, M_\perp$ and $\tan(\theta)$ over the maximum defined range. We then determine $c = c_1(M_{//,\infty}) c_2(M_\perp, \theta) = \frac{\ln(R)}{|M_{//,\infty}| + |M_\perp|\cot(\theta)}$. For $\theta = 90°$, $c$ is independent of $M_\perp$ and $\theta$ and we can plot the numerical solutions for $c_1$ as a function of $M_{//,\infty}$ assuming $c_2 = 1$. An expression $c_1(M_{//,\infty}) = a + b \cdot M_{//,\infty}^2$ fits these points (figure 8).

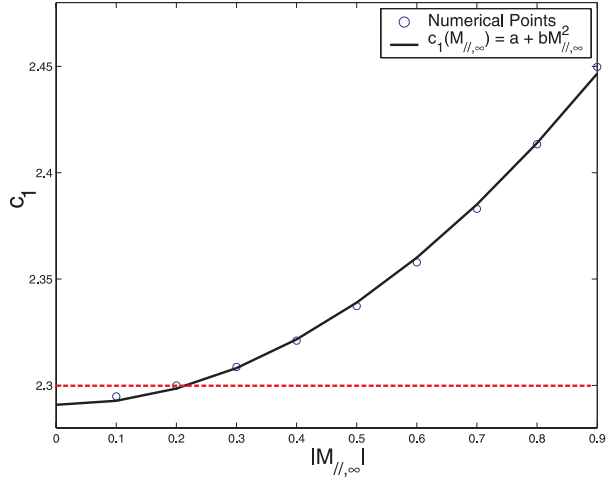
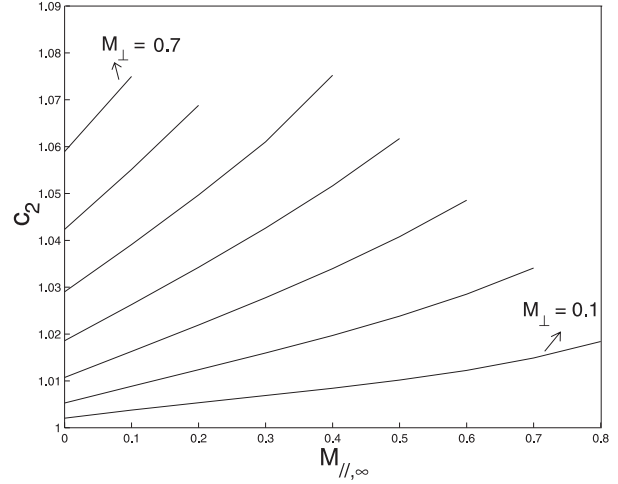

Figure 8: $c_1$ versus the parallel Mach number at the unperturbed plasma

Figure 9: $c_2$ versus $M_{//,\infty}$ for $\theta = 40°$ and $M_\perp = 0.1 \rightarrow 0.7$

To derive $c_2(M_\perp, \theta) = \dfrac{1}{a + b \cdot M_{//,\infty}^2} \cdot \dfrac{\ln(R)}{|M_{//,\infty}| + |M_\perp|\cot(\theta)}$ the following procedure is applied. We first keep the angle constant, for example $\theta_1 = 40°$, and vary both Mach numbers. We see that if we plot $c_2(M_\perp, \theta_1)$ as a function of $M_{//,\infty}$ (figure 9) the numerical solutions for all $|M_\perp| \leq 1$ can be fitted by the expression:

$$c_2(M_\perp, \theta_1) = e(M_\perp, \theta_1) + f(M_\perp, \theta_1).e^{M_\perp} \quad \text{with} \quad \begin{cases} e(M_\perp, \theta_1) = e_1(\theta_1) + e_2(\theta_1).e^{M_\perp} \\ f(M_\perp, \theta_1) = f_1(\theta_1) + f_2(\theta_1).e^{M_\perp} \end{cases}$$

We will now vary $0° \leq \theta \leq 180°$ to include its dependency. We found that the parameters $e_1$, $e_2$ and $f_1$, $f_2$ can be described by a common function $y(\theta) = p + q\theta^{-1}$ and write:

$$\begin{cases} e_i = e_{i,1} + e_{i,2}.\theta^{-1} \\ f_i = f_{i,1} + f_{i,2}.\theta^{-1} \end{cases} \quad (i = 1, 2)$$

If we now insert these definitions for $c_1$ and $c_2$ in our assumption for $c$ we get a non-linear expression of the form: $c(M_{//,\infty}, M_\perp, \theta) = a_1(M_{//,\infty})\left[a_2(M_\perp).\theta^{-1} + a_3(M_\perp)\right]$ with $a_i = a_{i1}.Z^2 + a_{i2}.Z + a_{i3}$ $(i = 1, 2, 3)$ for $\begin{cases} i = 1 \Rightarrow Z = M_{//,\infty} \\ i \neq 1 \Rightarrow Z = e^{M_\perp} \end{cases}$

The following table gives an overview of the values of the parameters that give the best fit.

| $a_{i,j}$ | 1 | 2 | 3 |
|---|---|---|---|
| 1 | 2.291 | 0 | 0.192 |
| 2 | 11.450 | -18.929 | 7.043 |
| 3 | -0.136 | 0.224 | 0.918 |

In this way a much better agreement with the exact numerical solutions over the complete parameter range, $0 \leq M_{//,\perp} \leq 1$ ; $0 \leq \theta \leq 180$, is achieved (figure 10).

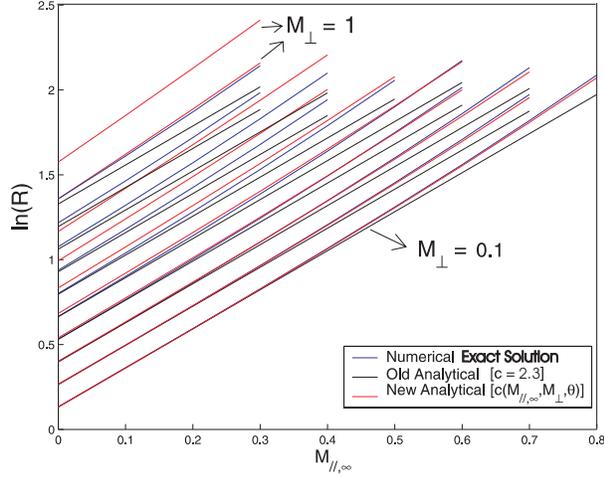 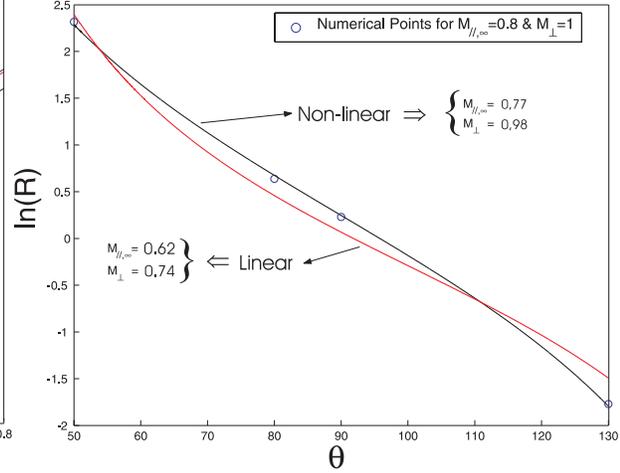

Figure 10: ln(R) versus $M_{//,\infty}$ for $\theta = 60°$ and $M_\perp = 0.1 \rightarrow 1$

Figure 11: The results of the 'old' linear and 'new' non-linear fit on the Mach numbers

Figure 10 shows that the improvement (for example when $\theta = 60°$) becomes important when higher flows exist. To quantify the effect, figure 11 shows an example of the values for the flows derived by fitting the old 'linear' and the new 'non-linear' function to four arbitrary data points (as in our experiments, four angles were available). The four input data for the least square fit are the numerical exact solutions for $M_{//} = 0.8$ and $M_\perp = 1$. The comparison of the two results from the linear and non-linear approach shows that, in the present case, the underestimation of the flows has been minimized. The underestimation of the perpendicular flow is reduced from 26% to 2%. The more precise value of the parallel Mach number becomes 0.77 instead the previous estimation of 0.62.

## 4. Conclusion

In this paper we formulated a new analytical expression for the factor $c(M_{//,\infty}, M_\perp, \theta)$ which takes into account the various dependencies of the parallel and perpendicular Mach numbers and the inclination angle of the collectors with respect to the magnetic field. This expression has been derived over the full parameter range $(-1 \leq M_{//,\perp} \leq 1$ and $0 \leq \theta \leq 180°)$. We showed that when a constant value for $c$ is chosen an error builds up when the flows grow and when the inclination angle deviates from 90°. During biasing experiments [4,5,6] higher flows are induced in the edge plasma and the use of the improved analytical approach is recommended.

Acknowledgement
This work was supported by FWO (Fund for Scientific Research-Flanders, Belgium).